\documentclass[journal]{IEEEtran}
\usepackage{hyperref}
\usepackage{cite}
\usepackage{bm}
\usepackage{float}
\usepackage{array}
\usepackage{multirow}
\usepackage[cmex10]{amsmath}
\usepackage{graphicx}
\usepackage{url}
\usepackage{color}
\usepackage{amssymb}
\usepackage{amsfonts}
\usepackage{amsmath}
\usepackage{extarrows}
\usepackage{graphicx}
\usepackage{amsfonts}
\usepackage{caption}
\usepackage{algorithm}
\usepackage{algpseudocode}
\usepackage{indentfirst}
\usepackage{caption}
\usepackage{esint} 
\usepackage{graphicx}
\usepackage{graphics}
\usepackage{pgfplots}
\usepackage{tikz}
\usepackage{epsfig}
\usepackage{soul}
\usepackage{caption}
\usepackage{gensymb}
\def\bA{{\mathbf{A}}}    
  \def\bH{{\mathbf{H}}} \def\bI{{\mathbf{I}}} 
  \def\bM{{\mathbf{M}}} \def\bN{{\mathbf{N}}} 
    
  \def\bW{{\mathbf{W}}}  \def\bY{{\mathbf{Y}}}

\def\ba{{\mathbf{a}}} \def\bb{{\mathbf{b}}}   
 \def\bg{{\mathbf{g}}} \def\bh{{\mathbf{h}}}  
   \def\bn{{\mathbf{n}}} 
    
    \def\by{{\mathbf{y}}}

\def\tcb{\textcolor{black}}

\begin{document}
\title{Learning to Estimate RIS-Aided mmWave Channels}
\author{Jiguang~He,~\IEEEmembership{Member,~IEEE,} Henk~Wymeersch,~\IEEEmembership{Senior Member,~IEEE,} Marco~Di~Renzo,~\IEEEmembership{Fellow,~IEEE}, and Markku~Juntti,~\IEEEmembership{Fellow,~IEEE}
\thanks{J. He and M. Juntti are with Centre for Wireless Communications, FI-90014, University of Oulu, Finland. H. Wymeersch is with Department of Electrical Engineering, Chalmers University of Technology, Gothenburg, Sweden. M. Di Renzo is with Universit\'e Paris-Saclay, CNRS, CentraleSup\'elec, Laboratoire des Signaux et Syst\`emes, 3 Rue Joliot-Curie, 91192 Gif-sur-Yvette, France.}
\thanks{This work is supported by Horizon 2020, European Union's Framework Programme for Research and Innovation, under grant agreement no. 871464 (ARIADNE). This work is also partially supported by the Academy of Finland 6Genesis Flagship (grant 318927), as well as the H2020 RISE-6G project under grant agreement number 101017011.}
}
 \maketitle
\begin{abstract}
Inspired by the remarkable learning and prediction performance of deep neural networks (DNNs), we apply one special type of DNN framework, known as model-driven deep unfolding neural network, to reconfigurable intelligent surface (RIS)-aided millimeter wave (mmWave) single-input multiple-output (SIMO) systems. We focus on uplink cascaded channel estimation, where known and fixed base station combining and RIS phase control matrices are considered for collecting observations. To boost the estimation performance and reduce the training overhead, the inherent channel sparsity of mmWave channels is leveraged in the deep unfolding method. It is verified that the proposed deep unfolding network architecture can outperform the least squares (LS) method with a relatively smaller training overhead and online computational complexity. 
\end{abstract}

\section{Introduction}
Reconfigurable intelligent surfaces (RISs) have recently been introduced for enhanced energy efficiency (EE), spectrum efficiency (SE), positioning accuracy, as well as network/physical-layer security~\cite{Wu2019, Liaskos18, di_renzo_smart_2019, wymeersch2019radio,Li2021}. The RIS, either being passive, active, or a hybrid combination of the former two, is used to smartly control the radio propagation environment, by virtue of multi-function capabilities, e.g., reflection, refraction, diffraction, scattering, and even absorption~\cite{Marco2020}. In the literature, the RIS is commonly used as an intelligent reflector, which breaks the well-known law of reflection~\cite{2019Basar}, to mitigate the blockage effect and expand the connectivity range, especially for millimeter wave (mmWave) communications. 

Since the RIS phase control and joint active and passive beamforming are sensitive to the channel state information (CSI) accuracy, the full potential of RIS cannot be achieved when the channel estimation (CE) is poorly performed. Therefore, accurate yet efficient CE methods for the individual channels  or the cascaded channel are of vital importance. In our previous works, we took advantage of the inherent channel sparsity and rank-deficiency features of the mmWave  multiple-input multiple-output (MIMO) channels and we applied the iterative reweighted method and the atomic norm minimization (ANM) method for estimating the channel parameters of RIS-aided mmWave MIMO systems~\cite{he2020,he2020anm}. These works fall into the category of conventional model-based approaches, which suit only for a small- or medium-sized RIS, base station (BS), and mobile station (MS).

As the number of RIS elements and BS/MS antennas continues to grow (this is an inevitable trend in the mmWave spectrum), more training overhead to obtain adequate CE performance, within the channel coherence time, via conventional model-based methods is required. Besides, the associated computational complexity will become inevitably high. These considerations motivate the application of data-driven or hybrid approaches for CE in RIS-aided communications~\cite{Zappone2019,Tekbiyik2021}. In~\cite{Elbir2020}, a convolutional neural network (CNN) was considered for RIS CE in a multi-user scenario. Therein, however, each user needs to first estimate its own channel and then map the estimate to the corresponding ground-truth channel, which naturally increases the computational complexity. Therefore, in this paper, we resort to a model-driven deep unfolding approach, which has already been used in MIMO detection and sparse signal recovery~\cite{Samuel2019, gregor2010learning}, for estimating the cascaded channel in RIS-aided mmWave single-input multiple-output (SIMO) systems. 


\tcb{Deep unfolding mimics the operation of conventional (projected) gradient descent algorithms, and it is capable of directly mapping the received pilot signals to the cascaded channel. Its computationally intensive training process can be executed offline and the online implementation/prediction phase only entails low-complexity calculations, e.g., matrix multiplications and additions, and element-wise operations. Besides, the step sizes and regularization parameters can be combined and optimized  during  the  training  of  the  deep  unfolding  model,  which  is  not  possible  in traditional gradient descent methods.} In this study, specifically, the rank-deficiency property of the cascaded channel is explicitly considered in the deep unfolding framework. It is verified that the deep unfolding scheme can outperform the least squares (LS) estimation \tcb{and the ANM methods~\cite{he2020anm}} in terms of normalized mean square error (NMSE) with a smaller training overhead and a reduced online computational complexity.  

\textit{Notations}: A bold lowercase letter $\ba$ denotes a vector, and a bold capital letter $\bA$ denotes a matrix. $(\cdot)^\mathsf{T}$ and $(\cdot)^\mathsf{H}$ denote the matrix or vector transpose and Hermitian transpose, respectively. $\mathrm{diag}(\ba)$ denotes a square diagonal matrix with the entries of $\ba$ on its diagonal, 
$\mathbb{E}[\cdot]$ is the expectation operator, $\mathbf{0}$ denotes the all-zero vector, $\bI_{M}$ denotes the $M\times M$ identity matrix, and $j = \sqrt{-1}$. $\|\cdot\|_\mathrm{F}$ and $\|\cdot\|_*$ denote the Frobenius norm and nuclear norm of a matrix, respectively, and $\|\cdot\|_2$ denotes the Euclidean norm of a vector.

\section{System Model}
\begin{figure}[t]
	\centering
\includegraphics[width=0.85\linewidth]{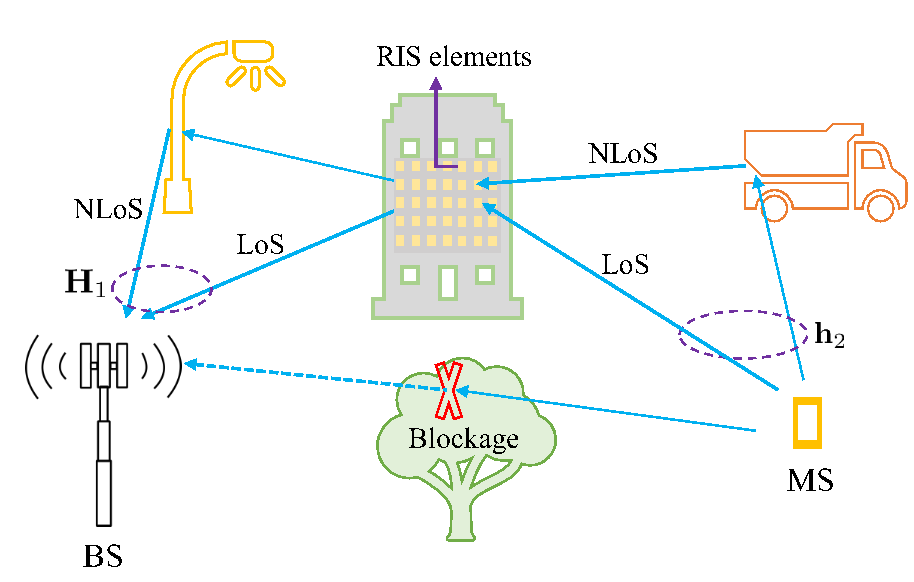}
	\caption{A typical scenario for maintaining the connectivity by deploying an RIS in a mmWave SIMO network.}
		\label{System_model}
		\vspace{-0.5cm}
\end{figure}
We consider a nearly-passive RIS-aided mmWave SIMO network, where the MS communicates with the BS via the RIS, as shown in Fig.~\ref{System_model}. The BS and RIS are equipped with multiple antennas and with nearly-passive scattering elements, respectively, while the MS is equipped with a single antenna. We further assume that the direct MS-BS channel is blocked.\footnote{When the direct MS-BS channel exists, we can estimate it by setting the RIS into an absorption mode. Then, we can estimate the cascaded channel by subtracting the direct MS-BS channel.} The RIS-BS channel, which is denoted as $\bH_1\in\mathbb{C}^{M \times N}$ with $M$ and $N$ being the number of antennas and scattering elements at the BS and RIS, respectively, can be written as
\begin{equation}\label{H_1}
\bH_1 = \sum\limits_{i = 1}^{L_1} g_{1,i} \boldsymbol{\alpha}(\phi_{1,i}) \boldsymbol{\alpha}^\mathsf{H}(\theta_{1,i}) = \bA(\boldsymbol{\phi}_1) \mathrm{diag}(\bg_1) \bA^\mathsf{H}(\boldsymbol{\theta}_1),
\end{equation}
 where $L_1 \ll \mathrm{min}\{M,N\}$ is the number of channel paths, including one line-of-sight (LoS) path (with $i = 1$) and multiple non-line-of-sight (NLoS) paths (with $i > 1$), and $g_{1,i}\in \mathbb{C}, $ $\theta_{1,i}\in \mathbb{R}$, and $\phi_{1,i}\in \mathbb{R}$ denote the propagation path gain, the angle of departure (AoD), and the angle of arrival (AoA) associated with the $i$th propagation path. The array response vector  $\boldsymbol{\alpha}(\phi_{1,i}) \triangleq [1, \;\;e^{ j \pi \sin(\phi_{1,i})}, \ldots,   e^{ j (M-1) \pi \sin(\phi_{1,i})}]^{\mathsf{T}} \in \mathbb{C}^{M\times 1}$ is obtained by assuming half-wavelength inter-element spacing, and  $\boldsymbol{\alpha}(\theta_{1,i})\in \mathbb{C}^{N \times 1}$ can be formulated in the same manner. We assume $g_{1,1}\sim\mathcal{CN}(0,\sigma_\text{LoS}^2)$, and define the vectors $\tilde{\bg}_1 \triangleq  [g_{1,2}, \ldots, g_{1, L_1}]^{\mathsf{T}} \sim \mathcal{CN}(\mathbf{0},\sigma_\text{NLoS}^2\bI_{L_1-1})$, $\bg_1 \triangleq [g_{1,1},\tilde{\bg}_1^{\mathsf{T}} ]^{\mathsf{T}}$, $\boldsymbol{\phi}_1 \triangleq [\phi_{1,1}, \ldots, \phi_{1,L_1}]^\mathsf{T}$, $\boldsymbol{\theta}_1 \triangleq [\theta_{1,1}, \ldots, \theta_{1,L_1}]^\mathsf{T}$,  $\bA(\boldsymbol{\phi}_1)  \triangleq [ \boldsymbol{\alpha}(\phi_{1,1}), \ldots,\boldsymbol{\alpha}(\phi_{1,L_1}) ]$, and $\bA(\boldsymbol{\theta}_1)  \triangleq [ \boldsymbol{\alpha}(\theta_{1,1}), \ldots, \boldsymbol{\alpha}(\theta_{1,L_1}) ]$. 

Similarly, the MS-RIS channel $\bh_2\in\mathbb{C}^{N \times 1}$ can be written as 
\begin{equation}
\bh_2 = \sum\limits_{i = 1}^{L_2} g_{2,i} \boldsymbol{\alpha}(\phi_{2,i}) =  \bA(\boldsymbol{\phi}_2) \bg_2.
\end{equation}
All the channel parameters in $\bh_2$ are defined as those in $\bH_1$. We also assume $g_{2,1}\sim\mathcal{CN}(0,\sigma_\text{LoS}^2)$, $\tilde{\bg}_2 \triangleq  [g_{2,2}, \ldots, g_{2, L_2}]^{\mathsf{T}} \sim \mathcal{CN}(\mathbf{0},\sigma_\text{NLoS}^2\bI_{L_2-1})$, and define $\bg_2 \triangleq [g_{2,1},\tilde{\bg}_2^{\mathsf{T}} ]^{\mathsf{T}}$ and $\bA(\boldsymbol{\phi}_2)  \triangleq [ \boldsymbol{\alpha}(\phi_{2,1}), \ldots,\boldsymbol{\alpha}(\phi_{2,L_2}) ]$ with $L_2 \ll N$.

The end-to-end uplink MS-RIS-BS channel (including the effect of the RIS) can be written as 
\begin{equation}\label{end2end_channel}
\bh =  \bH_1\boldsymbol{\Omega} \bh_2 = \bH_1\mathrm{diag}(\bh_2)\boldsymbol{\omega}, \end{equation}
where $\boldsymbol{\Omega} = \mathrm{diag}(\boldsymbol{\omega})$ is the RIS phase control matrix, with $\boldsymbol{\omega} = [\omega_1, \ldots, \omega_N]^\mathsf{T}$ and $|\omega_i| = 1$ for $\forall i$~\cite{Wu2019}. We are interested in low-cost and low-complexity implementations of RISs, we hence focus on RISs that can control only the phase response.  


In~\eqref{end2end_channel}, $\bH_1\mathrm{diag}(\bh_2)$ is referred to as the cascaded channel. By knowing it, we can optimize the RIS phase control matrix and BS beamforming/combining vector. Let us define it as $\bH_c\in\mathbb{C}^{M\times N}$, i.e.,
\begin{equation}\label{cascaded_channel}
   \bH_c = \bH_1\mathrm{diag}(\bh_2).
\end{equation}
Based on the considered assumptions, we have $\mathrm{rank}(\bH_1) = L_1$ and $\mathrm{rank}(\mathrm{diag}(\bh_2)) = N$. Thus $\mathrm{rank}(\bH_c)\leq \mathrm{min}\{\mathrm{rank}(\bH_1), \mathrm{rank}(\mathrm{diag}(\bh_2)) \} = L_1$.  The inherent channel sparsity (represented by the rank deficiency of the cascaded channel) can be applied in order to enable an efficient yet accurate CE of~\eqref{cascaded_channel}.

\section{Learning to Estimate}
In this section, we first introduce the channel sounding procedure, and then describe the optimization problem formulation for recovering the cascaded channel by using conventional (model-based) methods. Finally, we describe the model-driven deep unfolding method for estimating the rank-deficient cascaded channel.  

\subsection{Channel Sounding}
During the sounding process, pilot signals are sent from the MS to the BS via the RIS. A different RIS phase control matrix is considered for each channel use while the combining matrix at the BS is fixed. The received signal at channel use $k$, for $k = 1,\ldots, K$, can be written as
\begin{equation} \label{rec_sig_BS}
    \by[k] = \bW^\mathsf{H}[k] \bH_1 \boldsymbol{\Omega}[k] \bh_2 s[k] + \bW^\mathsf{H}[k] \bn[k],
\end{equation}
where $\bW[k] \in \mathbb{C}^{M \times N_\text{W}}$ is the combining matrix at the BS with $N_\text{W}$ denoting the number of columns,\footnote{We consider an analog combining matrix, which is a suitable choice for fulfilling the requirements of reduced-complexity hybrid precoding architectures commonly assumed for mmWave MIMO transceivers. When $N_\text{W} > N_\text{RF}$, with $N_\text{RF}$ being the number of radio frequency (RF) chains at the BS, we need $K \lceil N_\text{W}/N_\text{RF}\rceil$ channel uses to complete the sounding process. Otherwise, $K$ channel uses are sufficient.} $s[k]$ is the pilot signal sent by the MS, and $\bn[k] \sim \mathcal{CN}(0, \sigma^2)$ is the additive white Gaussian noise (AWGN) at the BS. 

The received signal $\by[k]$ in~\eqref{rec_sig_BS} can be reformulated as 
\begin{equation}
    \by[k] = \bW^\mathsf{H}[k] \bH_c \boldsymbol{\omega}[k] s[k] + \bW^\mathsf{H}[k] \bn[k],
\end{equation}
where $\boldsymbol{\Omega}[k] = \mathrm{diag}(\boldsymbol{\omega}[k])$. 

Without loss of generality, we assume $s[1] = s[K] =1$ and $\bW[1] = \bW[K] = \bW$. The received signals $\bY = [ \by[1], \ldots, \by[K]]$ can be rewritten as
\begin{equation}\label{Rec_signals}
    \bY = \bW^\mathsf{H} \bH_c \bar{\boldsymbol{\Omega}} + \bW^\mathsf{H}\bN,
\end{equation}
where $\bar{\boldsymbol{\Omega}} = [\boldsymbol{\omega}[1], \ldots, \boldsymbol{\omega}[K]]$ and $\bN = [\bn[1],\ldots, \bn[K]]$. An additional vectorization step is considered for all the terms in~\eqref{Rec_signals}, resulting in 
\begin{equation}\label{vec_rec_signal}
    \by = (\bar{\boldsymbol{\Omega}}^\mathsf{T} \otimes \bW^\mathsf{H}) \bh_c + \bn,
\end{equation}
where $\by = \mathrm{vec}(\bY), \bh_c = \mathrm{vec}(\bH_c)$, and $\bn = \mathrm{vec}(\bW^\mathsf{H}\bN)$. Based on the vectorized received signal $\by$, we need to estimate the vectorized cascaded channel $\bh_c$. Let us define $\boldsymbol{\Psi} = \bar{\boldsymbol{\Omega}}^\mathsf{T} \otimes \bW^\mathsf{H}$. Then, \eqref{Rec_signals} can be simplified as $\by = \boldsymbol{\Psi} \bh_c + \bn$. Based on the obtained signal model, the objective of this paper is to extract $\bh_c$ from the noisy received signal $\by$ by assuming that the matrix $\boldsymbol{\Psi}$ is known. This is accomplished by using the deep unfolding method, which is detailed next. 

\subsection{Optimization Problem Formulation}
In order to recover $\bh_c$ from the noisy observation $\by$, we formulate the following regularized optimization problem
\begin{equation}\label{Optization_problem1}
    \hat{\bh}_c = \arg\min_{\bh_c} \|\by - \boldsymbol{\Psi}\bh_c\|_2^2 + \lambda \mathrm{rank}(\bH_c),
\end{equation}
which takes into consideration the rank deficiency of the cascaded channel $\bH_c$ and the impact of noise~\cite{Chen2018}. The regularization parameter $\lambda > 0$ is introduced to control the tradeoff between the data fitting and the rank of the cascaded channel. The optimization problem in~\eqref{Optization_problem1} can be further reformulated as 
\begin{equation}\label{Optization_problem2}
    \hat{\bh}_c = \arg\min_{\bh_c} \|\by - \boldsymbol{\Psi}\bh_c\|_2^2 + \lambda \|\bH_c\|_*,
\end{equation}
which is obtained by relaxing $\mathrm{rank}(\bH_c)$ with its nuclear norm, i.e., $\|\bH_c\|_*$. This is a convenient approach because $\mathrm{rank}(\bH_c)$ is a noncovex function of $\bH_c$. In addition, $\|\bH_c\|_{\mathrm{F}} \leq \|\bH_c\|_* \leq \sqrt{r} \|\bH_c\|_{\mathrm{F}}$ with $r \geq 1$. Notably, when $\mathrm{rank}(\bH_c) =1$ (e.g., $\bH_1$ has only the LoS path), we have $\|\bH_c\|_* = \|\bH_c\|_{\mathrm{F}}$. Also, the singular values of $\bH_c$ have a high probability to fulfill the following condition: $\sigma_1 \gg \sigma_2 > \cdots > \sigma_{L_1}$ with $\sigma_i$ being the $i$th largest singular value of $\bH_c$, which results in $\|\bH_c\|_* \approx \|\bH_c\|_{\mathrm{F}}$. Thus, we further replace $\mathrm{rank}(\bH_c)$ in~\eqref{Optization_problem2} with $\|\bH_c\|_{\mathrm{F}}$, i.e., $\|\bh_c\|_2$, which yields
\begin{equation}\label{Optization_problem3}
    \hat{\bh}_c = \arg\min_{\bh_c} \|\by - \boldsymbol{\Psi}\bh_c\|_2^2 + \lambda \|\bh_c\|_2.
\end{equation}
To accurately solve this optimization problem, we need to carefully choose the regularization parameter $\lambda$. The optimal value of $\lambda$ is, however, difficult to obtain. As a reference, $\lambda$ is chosen equal to $4 \sigma^2 \sqrt{\frac{MN(M+N)\log(M+N)}{N_\text{W} K}}$~\cite{Chen2018}. 
\subsection{Model-Driven Deep Unfolding}
Deep unfolding is a deep neural network framework that mimics the conventional gradient descent method. The difference lies in that deep unfolding is able to learn from a large amount of (synthetic) data with enhanced performance and reduced online implementation complexity and number of iterations. Typically, this is exemplified in a reduced number of layers in the deep unfolding
network. When solving the optimization problem in~\eqref{Optization_problem3} by using the (conventional) gradient descent method, we iteratively update  $\bh_c^{(i)}$, with $(i)$ denoting the iteration index, as follows 
\begin{equation}\label{GD_iteration}
    \bh_c^{(i)} = \bh_c^{(i-1)}- \beta  \nabla f(\bh_c^{(i-1)}), 
\end{equation}
where $0<\beta<1$ is the step size and $\nabla f(\bh_c^{(i-1)})$ is the gradient of $f(\bh_c) = \|\by - \boldsymbol{\Psi}\bh_c\|_2^2 + \lambda  \|\bh_c\|_2$ evaluated at $\bh_c^{(i-1)}$, which can be expressed as 
\begin{equation} \label{gradient}
    \nabla f(\bh_c^{(i-1)}) = (\boldsymbol{\Psi}^\mathsf{H} \boldsymbol{\Psi}  \bh_c^{(i-1)} -\boldsymbol{\Psi}^\mathsf{H} \by ) + \lambda  \bh_c^{(i-1)}/\|\bh_c^{(i-1)}\|_2, 
\end{equation}
when $\lambda$ is fixed. The initial value, i.e, $\bh_c^{(0)}$, can be set equal to the all-zero vector. In this case, the denominator $\|\bh_c^{(i-1)}\|_2$ in the last term of~\eqref{gradient} needs to be modified as $\|\bh_c^{(i-1)}\|_2 + \epsilon$ with $\epsilon >0$ when $i = 1$. 

Substituting~\eqref{gradient} into~\eqref{GD_iteration}, we obtain 
\begin{equation}\label{GD_iteration1}
    \bh_c^{(i)} = \bh_c^{(i-1)}- \beta \boldsymbol{\Psi}^\mathsf{H} \boldsymbol{\Psi}  \bh_c^{(i-1)} +\beta \boldsymbol{\Psi}^\mathsf{H} \by  - \beta  \lambda  \bh_c^{(i-1)}/\|\bh_c^{(i-1)}\|_2.
\end{equation}
The Gram matrix $\boldsymbol{\Psi}^\mathsf{H} \boldsymbol{\Psi}$, the compressed statistics $\boldsymbol{\Psi}^\mathsf{H}\by$, and $\bh_c^{(i-1)}$, are needed to apply the gradient descent algorithm. Therefore, these three terms constitute the input variables of the deep unfolding neural network model. 

\begin{figure}[t]
	\centering
\includegraphics[width=0.85\linewidth]{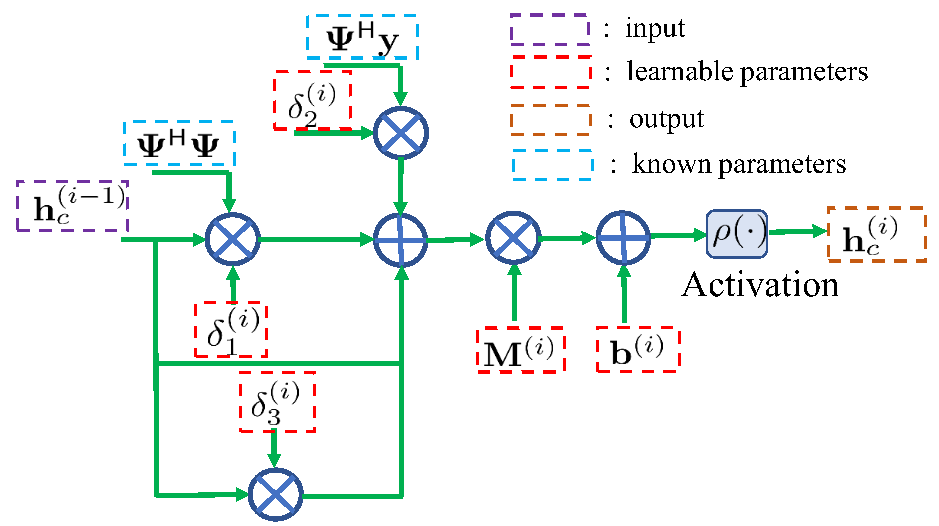}
	\caption{The $i$th layer of the deep unfolding network model for estimating the cascaded channel vector.}
		\label{Deep_unfolding_one_layer}
		\vspace{-0.5cm}
\end{figure}

To be specific, the $i$th layer of the deep unfolding model for mimicking the gradient descent iteration in~\eqref{GD_iteration1} is introduced in Fig.~\ref{Deep_unfolding_one_layer}. The accuracy and convergence speed of the channel estimate in~\eqref{GD_iteration1} highly depends on the specific choice of the step size $\beta$ and the regularization parameter $\lambda$. In the considered deep unfolding model, these parameters are learnable parameters that are automatically determined during the data-driven training phase. To this end,
we introduce three generalized learnable parameters 
$\delta_{1}^{(i)} \in [-1, \;0]$, $\delta_{2}^{(i)}\in [0, \;1]$, and $\delta_{3}^{(i)}\in [-1, \;0]$, for the $i$th layer of the deep unfolding model in Fig.~\ref{Deep_unfolding_one_layer}. 
In~\eqref{GD_iteration1}, more specifically, $-\beta$ and $\beta$ in the second and third term are replaced by  $\delta_{1}^{(i)}$ and $\delta_{2}^{(i)}$, respectively, and the product $-\beta \lambda /\|\bh_c^{(i-1)}\|_2$ in the last term is unfolded in the learnable parameter $\delta_{3}^{(i)}$. As shown in Fig.~\ref{Deep_unfolding_one_layer}, 
in order to further enhance the prediction capabilities of the deep unfolding network model, we serially concatenate the term $\bh_c^{(i-1)}+ \delta_{1}^{(i)} \boldsymbol{\Psi}^\mathsf{H} \boldsymbol{\Psi}  \bh_c^{(i-1)} +\delta_{2}^{(i)} \boldsymbol{\Psi}^\mathsf{H} \by  + \delta_{3}^{(i)}  \bh_c^{(i-1)}$ with a learnable weight matrix $\bM^{(i)}$, a bias vector $\bb^{(i)}$, and a non-linear activation function. 

\begin{figure}[t]
	\centering
\includegraphics[width=0.85\linewidth]{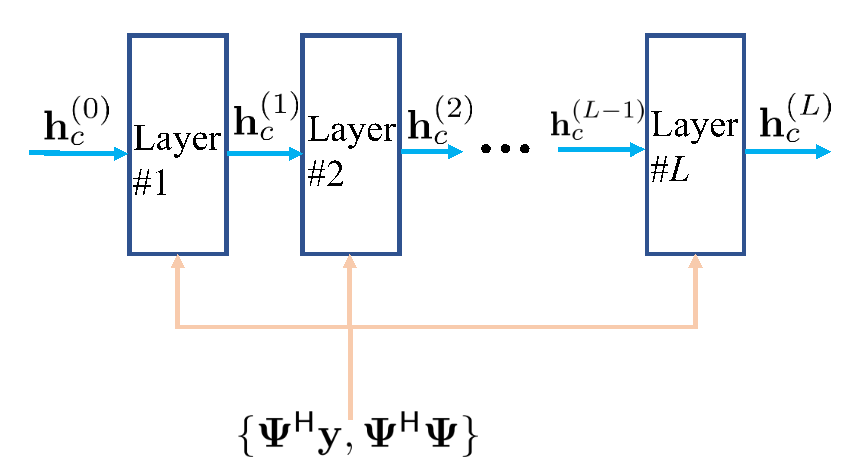}
	\caption{Complete deep unfolding network model for channel estimation, which comprises $L$ layers from Fig.~\ref{Deep_unfolding_one_layer}}
		\label{Deep_unfolding_framework}
		\vspace{-0.5cm}
\end{figure}

The complete deep unfolding network model is illustrated in Fig.~\ref{Deep_unfolding_framework} and it comprises $L$ layers from Fig.~\ref{Deep_unfolding_one_layer}. In particular, the observations
$\{\boldsymbol{\Psi}^\mathsf{H} \boldsymbol{\Psi},\boldsymbol{\Psi}^\mathsf{H}\by \}$ are input to all the layers of the deep unfolding network model. \tcb{The online computational complexity of the proposed scheme is $\mathcal{O}(M^2N^2L)$, which is smaller than the ANM based scheme that requires $\mathcal{O}((M+N)^6)$ per iteration~\cite{he2020anm}.}


\section{Numerical Results}
In this section, we evaluate the performance of the proposed deep unfolding network model against two benchmark schemes: (i) the LS estimator and (ii) the direct solution of~\tcb{\eqref{Optization_problem3}} by using the CVX toolbox.\footnote{The scripts of the implementation are available at \url{https://github.com/jiguanghe/RISCE}.} We evaluate the impact of the training overhead, the training SNR, the number of paths, and the angular parameter distribution. As far as the RIS phase control matrices as concerned, their diagonal elements are set equal to the column vectors of a discrete Fourier transform (DFT) matrix. A set of orthonormal vectors are considered for $\bW$, e.g., the normalized column vectors from a DFT matrix. It is worth mentioning that we transform the data from the complex-valued domain to the real-valued domain before applying the deep unfolding network model. The channel $\bh_c^{(0)}$ is set equal to the all-zero vector. In the first $L-1$ layers, we use \textit{relu} activation functions, while no activation function is applied in the last layer. The loss function during the training phase is the NMSE between the output cascaded channel vector and the ground-truth cascaded channel vector. The \textit{Adam} algorithm is used for training, whose learning rate is $0.001$ during the first $20$ epochs and it is halved in the remaining epochs, and the batch size is $64$. We use $1e5$ samples for training and $1e4$ samples for testing. The parameter setup is summarized in Table~\ref{tab1:parameter}.

\subsection{Impact of the Training Overhead}
\begin{figure}[t]
	\centering
\includegraphics[width=0.85\linewidth]{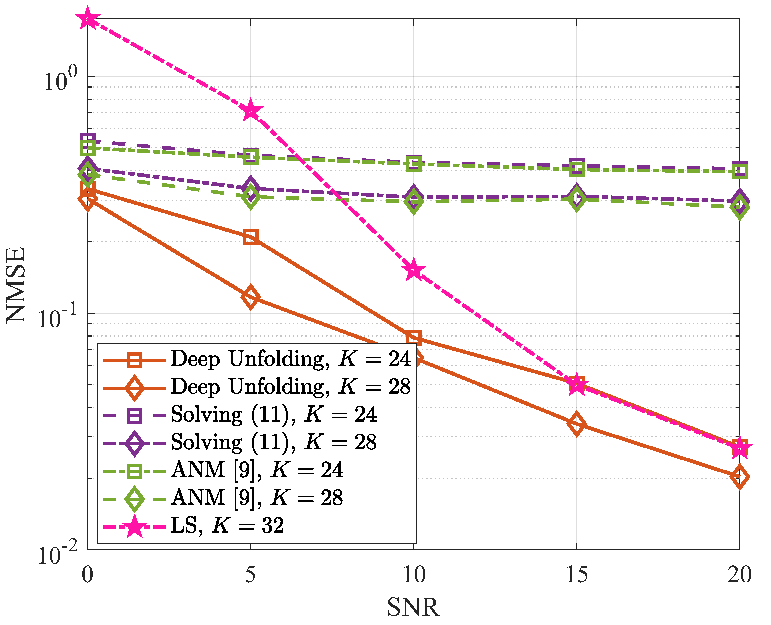}
	\caption{\tcb{Impact of training overhead on estimating the cascaded channel for a  $1\times 16$ SIMO system with $N=32$ and $L_1=L_2=1$. Deep unfolding vs.\ LS estimation, ANM~\cite{he2020anm}, and the direct solution of~\eqref{Optization_problem3}.}}
		\label{LS_MIMO16x1_RIS32_training_overhead}
		\vspace{-0.5cm}
\end{figure}

Fig.~\ref{LS_MIMO16x1_RIS32_training_overhead} shows the NMSE for a $1\times 16$ SIMO system (i.e., $M =16$) by using the proposed deep unfolding approach as a function of the training overhead. During the training phase, the SNR is $\gamma= 1/\sigma^2 = 20$ dB. From Fig.~\ref{LS_MIMO16x1_RIS32_training_overhead}, we see that the proposed scheme with $K=24$ channel uses for training outperforms the LS estimator even if the latter uses a longer training phase with $K=32$.
Also, as expected, the higher the training overhead is, the lower the NMSE of the proposed scheme is. It is worth noting that the proposed deep unfolding method outperforms the numerical solution of~\tcb{\eqref{Optization_problem3}} by using CVX, \tcb{and the ANM algorithm}. This is attributed to the learning capability of the deep unfolded network through the learnable parameters introduced in Fig.~\ref{Deep_unfolding_one_layer}.\footnote{\tcb{If the regularization parameter $\lambda$ in  \eqref{Optization_problem3} can be optimally designed, better performance can be expected.}}


\begin{table}[t]
    \centering
   \caption{\tcb{Parameter Setup.}}
    \label{tab1:parameter}
    \begin{tabular}{cc||cc}
        \hline
        Parameter  & Value & Parameter  & Value \\
       \hline
        $L_1$ & $\{1,2,3\}$ &$\bh_c^{(0)}$&$\mathbf{0}$ \\
        $L_2$ & $\{1,2,3\}$ & Optimizer& \textit{Adam}\\
        $\sigma_\text{LoS}^2$&$1$ & Learning rate & $0.001, 0.0005$\\
        $\sigma_\text{NLoS}^2$&$0.01$& Batch size&$64$\\
        $M$  &  $16$ & $\rho(\cdot)$& \textit{relu} \\ 
        $N$ & $32$ & Training samples& $1e5$ \\ 
        $L$  &  $15$ & Testing samples& $1e4$\\
        $\sin(\boldsymbol{\theta}_1)$ & $\mathcal{U}[0,\; 1]$ 
        & Loss& NMSE \\ $\sin(\boldsymbol{\phi}_1)$& $\mathcal{U}[0,\; 1]$ &$\sin(\boldsymbol{\phi}_2)$& $\mathcal{U}[0,\; 1]$ \\
        $N_\text{W}$ & $8$&
        $K$ & $\{24, 28\}$ \\
       \hline
    \end{tabular}
\end{table}

\subsection{Impact of the Number of Paths}
\begin{figure}[t]
	\centering
\includegraphics[width=0.85\linewidth]{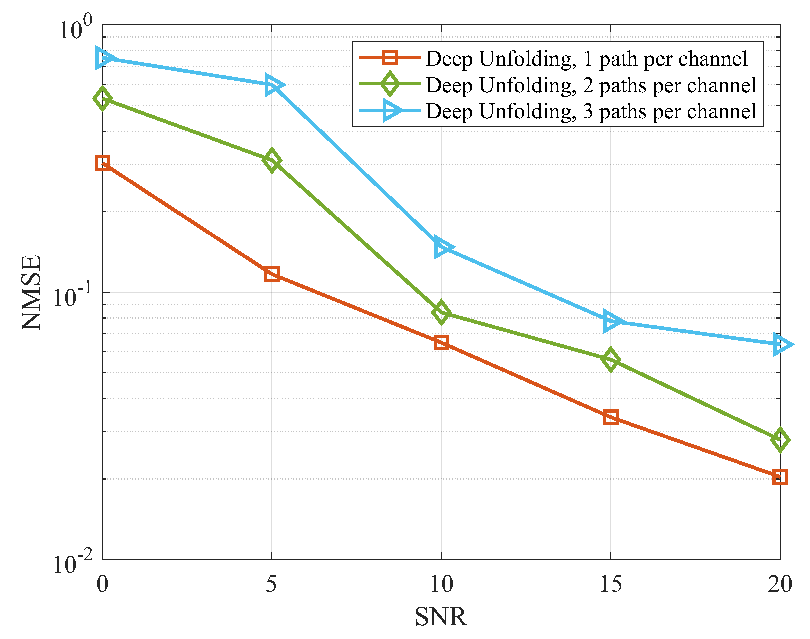}
	\caption{Impact of the number of paths on estimating the cascaded channel for a $1\times 16$ SIMO system with $N=32$.}
		\label{LS_MIMO16x1_RIS32_number_of_paths}
			\vspace{-0.5cm}
\end{figure}

In this subsection, we study the impact of the number of paths, which is increased from one to two and three. The SNR for training is $\gamma = 20$ dB and $K=28$. The corresponding NMSE is shown in Fig.~\ref{LS_MIMO16x1_RIS32_number_of_paths}. When the number of paths increases, the rank of the cascaded channel increases accordingly. In this case, Fig.~\ref{LS_MIMO16x1_RIS32_number_of_paths} shows that the NMSE increases as the number of paths increases. In other words, by keeping the training overhead fixed, the proposed deep unfolding method benefits from the sparsity of the channel, i.e., the number of paths is small. 
\subsection{Impact of the Training SNR}
\begin{figure}[t]
	\centering
\includegraphics[width=0.85\linewidth]{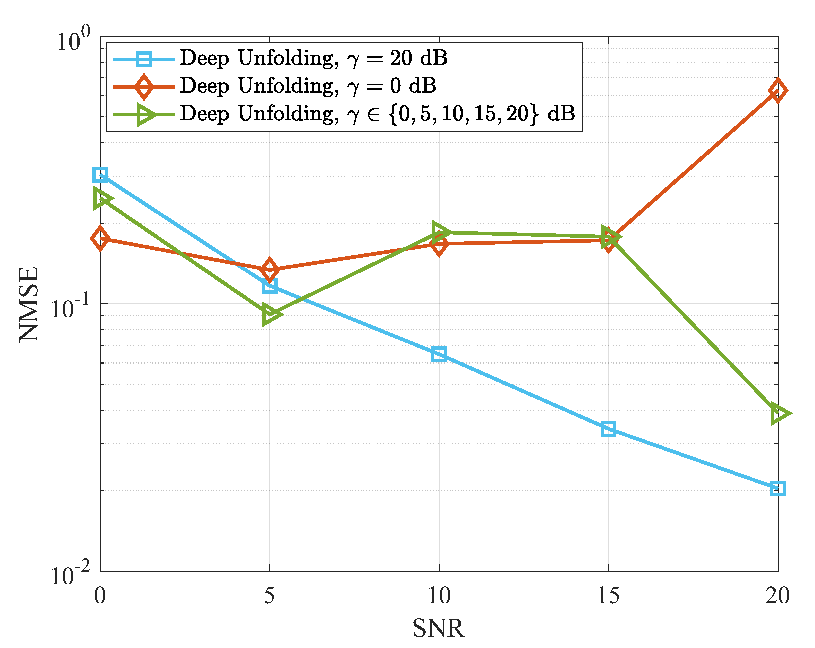}
	\caption{Impact of the training SNR on estimating the cascaded channel for a $1\times 16$ SIMO system with $N = 32$.}
		\label{LS_MIMO16x1_RIS32_training_SNR}
			\vspace{-0.5cm}
\end{figure}

In this subsection, we study the impact of the training SNR on the estimation performance. Two SNR values, i.e., $\gamma = 0$ dB, $\gamma = 20$ dB, and an SNR that varies in the set $\gamma \in \{0, 5,10,15,20\}$ dB are considered. The corresponding NMSE is shown in Fig.~\ref{LS_MIMO16x1_RIS32_training_SNR}. We observe that the proposed deep unfolding method provides the best NMSE when it is trained at a high SNR except if the operating SNR during the test phase is too low. \tcb{In other words, nearly noise-free training samples bring the best performance in the high SNR regime, and vice versa, as depicted in Fig.~\ref{LS_MIMO16x1_RIS32_training_SNR}. The training SNR plays a critical role in the prediction performance of the deep unfolding models.\footnote{\tcb{If the SNR value(s) can be incorporated into the proposed deep unfolding structure, better performance is expected, which is left for our future investigation.}}}

\subsection{Impact of the Angular Parameter Distribution}
\begin{figure}[t]
    \centering
 \includegraphics[width=0.85\linewidth]{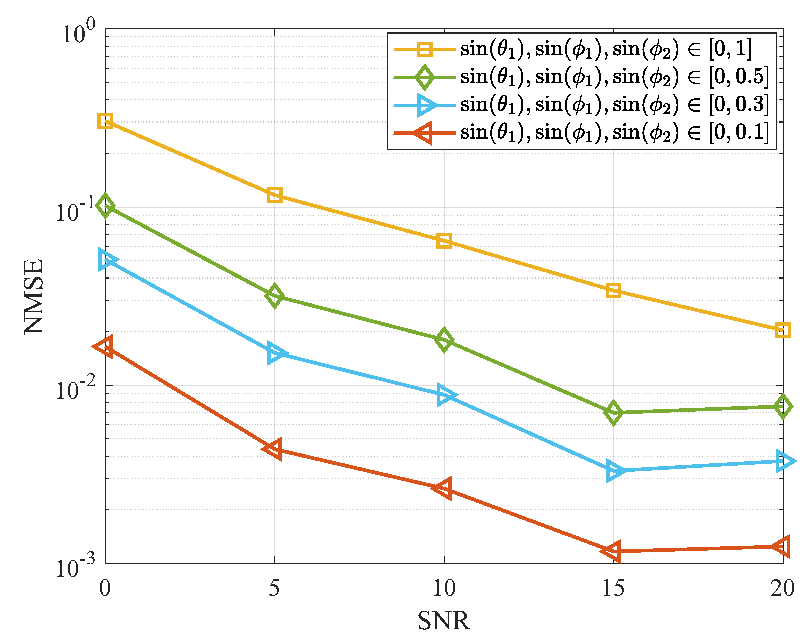}
	\caption{Impact of the angular parameter distribution on estimating the cascaded channel for a $1\times 16$ SIMO system with $N=32$.}
		\label{LS_MIMO16x1_RIS32_parameter_distribution}
			\vspace{-0.5cm}
\end{figure}

\tcb{Unlike the previous study in which the angular parameters are distributed as $\mathcal{U}[0,\; 1]$,} in this subsection, we evaluate the impact of the angular parameter distribution when estimating the cascaded channel. The corresponding NMSE is shown in Fig.~\ref{LS_MIMO16x1_RIS32_parameter_distribution}. We observe that the NMSE decreases when the range of $\sin(\boldsymbol{\theta}_1), \sin(\boldsymbol{\phi}_1), \sin(\boldsymbol{\phi}_2)$ decreases. In other words, the estimation accuracy of the proposed deep unfolding method increases when the individual channels are subject to a reduced variability.

\vspace{-0.5cm}
\section{Conclusions}
In this letter, we have introduced a deep unfolding model for efficiently estimating the end-to-end RIS channel in mmWave SIMO systems. With the aid of simulation results, we have shown that the proposed approach can outperform three benchmark schemes based on the LS method, a CVX-based numerical solution of the channel estimation problem, \tcb{and the ANM algorithm}. The impact of the number of paths, the training SNR, and the angular parameter distribution on the estimation accuracy has been investigated. In addition, the proposed deep unfolding network model has a low online prediction complexity, since it requires the computation of vector matrix multiplications and additions. On the other hand, the LS estimation methods usually require a matrix inversion.

Possible generalizations of the present work include the channel estimation problem in RIS-aided multiuser MIMO systems, and the joint optimization of the active and passive beamforming at the BS and at the RIS, respectively.

\bibliographystyle{IEEEtran}
\bibliography{IEEEabrv,Ref}

\end{document}